\pdfoutput=1

\documentclass[11pt]{article}

\usepackage[]{ACL2023}

\usepackage{times}
\usepackage{latexsym}
\usepackage{graphicx}
\usepackage{multirow}
\usepackage{subfiles}
\usepackage{url}
\usepackage{amsmath}
\usepackage{latexsym}
\usepackage{xcolor}
\usepackage{float}
\usepackage{subcaption}
\usepackage{array}
\usepackage{caption}

\usepackage[T1]{fontenc}

\usepackage[utf8]{inputenc}

\usepackage{microtype}

\usepackage{inconsolata}

\title{ReXCL: A Tool for Requirement Document Extraction and Classification}

 \author{Paheli Bhattacharya, Manojit Chakraborty \\  {\bf Santhosh Kumar Arumugam} \and {\bf Rishabh Gupta} \\
         Bosch Research and Technology Centre, Bangalore, India\\
         \texttt{\{paheli.bhattacharya, manojit.chakraborty\}@in.bosch.com},\\ 
         \texttt{\{{santhoshkumar.arumugam, gupta.rishabh\}@in.bosch.com}}}

\begin{document}
\maketitle
\begin{abstract}
This paper presents the ReXCL tool, which automates the extraction and classification processes in requirement engineering, enhancing the software development lifecycle. The tool features two main modules: Extraction, which processes raw requirement documents into a predefined schema using heuristics and predictive modeling, and Classification, which assigns class labels to requirements using adaptive fine-tuning of encoder-based models. The final output can be exported to external requirement engineering tools. Performance evaluations indicate that ReXCL significantly improves efficiency and accuracy in managing requirements, marking a novel approach to automating the schematization of semi-structured requirement documents.
\end{abstract}

\section{Introduction}
Extraction and classification are vital activities in requirement engineering that ensure the effective gathering, organization, and management of the requirements for a software project to ensure that the software meets the needs of all stakeholders~\cite{ser1}. Currently, the extraction activities are largely manual which requires processing of multiple formats of requirement documents and then mapped in a defined schema. Whereas, the classification activity takes the schematized version of the requirement texts and classify them into functional, non-functional, etc. categories to facilitate management and analysis. This is again time-consuming and human intensive activity which have inconsistencies, confusions and quality issues. This calls for the need of AI systems that can assist requirement engineers in the process~\cite{rajbhoj2024accelerating}.

In this study, we introduce the ReXCL tool, designed to automate the extraction and classification processes in requirement engineering, thereby significantly improving the software development lifecycle. 
The tool, illustrated in Figure~\ref{fig:overall_workflow}, consists of two main modules -- Extraction and Classification. Given a input document, the Extraction module processes the raw document to conform to a predefined schema. This module utilizes a mix of heuristics and predictive modeling techniques (ref. Section~\ref{sec:extraction}). The schematized document is then passed on to the Classification module for assigning class labels to each requirement text (row) into one of four classes -- \textit{Info, Header, Functional Requirement} and \textit{Non-Functional Requirement}. This module uses adapative fine-tuning of encoder-based models like BERT~\cite{devlin2019bertpretrainingdeepbidirectional} (ref. Section 2.2). The final output can then be downloaded and exported to external RE tools like IBM Doors, Jira, etc. We analyse the performance of the modules through both automatic and human evaluation. The results suggest the tool's impressive performance in both modules (ref. Section~\ref{sec:eval}). While there has been prior work on headline detection in documents~\cite{heading}, to the best of our knowledge, this is the first work that attempts to automatically schematize semi-structured requirement documents in a holistic manner.

\section{ReXCL System}
\begin{figure*}[t]
\caption{The ReXCL tool pipeline. The input is a customer requirement. The extraction module parses the document to produce a structured tabular output. The classification module then classifies each requirement text (row). The final output can then be exported to the tools like IBM Doors.}
\label{fig:overall_workflow}
\centering
\includegraphics[width = \linewidth]{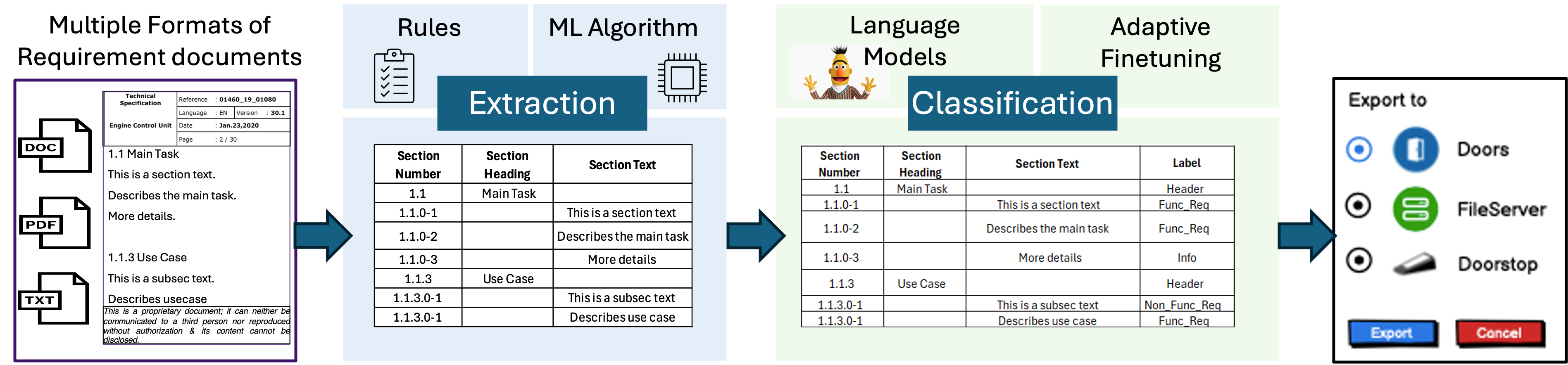}
\end{figure*}

The ReXCL tool, depicted in Figure~\ref{fig:overall_workflow}, comprises of two primary modules: Extraction and Classification. When provided with an input document, the Extraction module parses the raw document to extract the \textbf{section numbers} (e.g. 1.1, 1.1.3), the corresponding \textbf{section headings} (e.g. Main Task, Use Case) and the associated \textbf{section texts} for each section $i$. The components are merged into a structured format to produce the final extraction output of the input document as given in Equation 1: 
\begin{equation}
\begin{aligned}
    extraction = \bigcup_{i=1}^{n}< number_i, heading_i,\\
    [text_{i_{1}}, text_{i_{2}} ... , text_{i_{m}}]>
    \end{aligned}
\end{equation}

We refer to $\mathbf{section\ title_i}$ as a combination of section $number_i$ and $heading_i$ (e.g. 1.1 Main Task), $m$ denotes the number of section texts associated with a section $title_i$ and $n$ denotes the number of section $titles$ present in the document. The structured document $extract$ is subsequently forwarded to the classification module.

The Classification module is responsible for classifying each $i$ in $extract$ into one of four classes -- \textit{Info, Header, Functional Requirement} and \textit{Non-Functional Requirement}. The final output then consists of $final\_output$ (as given in Equation 2) in a structured format. This data can then be exported to requirement engineering tools in \textit{excel}, \textit{csv} or \textit{json} formats. 
\begin{equation}
    \begin{aligned}
final\_output= \bigcup_{i=1}^{n}< number_i, heading_i,\\
    text_{i_{1}}, text_{i_{2}} ... , text_{i_{m}}], class_i
    \end{aligned}
\end{equation}

\subsection{Extraction}
In this section we describe the extraction module. Broadly, the idea is to identify textual units (lines, paragraphs, etc.) that are section titles (consisting of section number and section heading). From the section title, one can parse out the section number (consisting of digits) and the section heading as shown in Figure~\ref{fig:ext}. Textual units between any two section titles is a section text and can be paired with section title appearing before it.

$\bullet$ \textbf{Intermediate Representation}: The input requirement document can be either in \textit{.pdf}, \textit{.doc} or \textit{.txt}. We first convert the document into an intermediate text representation, which produces sentences. We then classify each line as a section title or a section text. In this work, we explore the following $2$ approaches to generate these representations:

\textbf{(i)} Using regular text conversion packages like PyMuPDF~\footnote{\url{https://pypi.org/project/PyMuPDF/}}, PyPDF2~\footnote{\url{https://pypi.org/project/PyPDF2/}} that convert all the contents into text: We observe that the outputs from these packages do not automatically distinguish between section titles and section texts. So a rule-based parser, that detects patterns of the form ($digit<dot>digit<dot>digit<dot>....<dot>digit\ Text $) needs to be applied, with the assumption that section titles will start with a digit followed by the text of the heading (e.g. 1.4 Requirements) and can be summarized by the pattern above. The drawbacks of this approach are: (a)~this pattern may be wrongly classify section texts as titles since it may be present in a section text also (b)~the section numbers may appear as roman numerals (e.g. IV, VII) or alphabets (e.g. A, B) and several rules need to be handcrafted (c)~the tables are flattened into lines of text, thus loosing its structural identity.(d)~text styles are not preserved.

\textbf{(ii)} Using markdown text conversion packages like PyMuPDF4LLM~\footnote{\url{https://pymupdf.readthedocs.io/en/latest/pymupdf4llm/}} : To alleviate the problem of having a rule-based parser that detects section titles based on patterns, methods that automatically detect such patterns, in an unsupervised manner, will be beneficial. We find that converting data to markdown text achieves this while preserving several features -- (a)~the section titles are marked with "\#" making it easier to detect them. This alleviates the problem of rule-based pattern matching (b)~the table structure is preserved with "|" separating the columns (e.g. |col1|col2|col3|) (c)~text styles, e.g. \textbf{bold} gets converted to $**$bold$**$ which helps in better output rendering.

\begin{figure*}[th]
\caption{The extraction module workflow; the input is a raw document, and the output is a final structured output containing section number, section heading and section text. The components used are intermediate text representation, header-footer removal, section information extraction and final output generation.}
\label{fig:extraction_workflow}
\centering
\includegraphics[width=\linewidth]{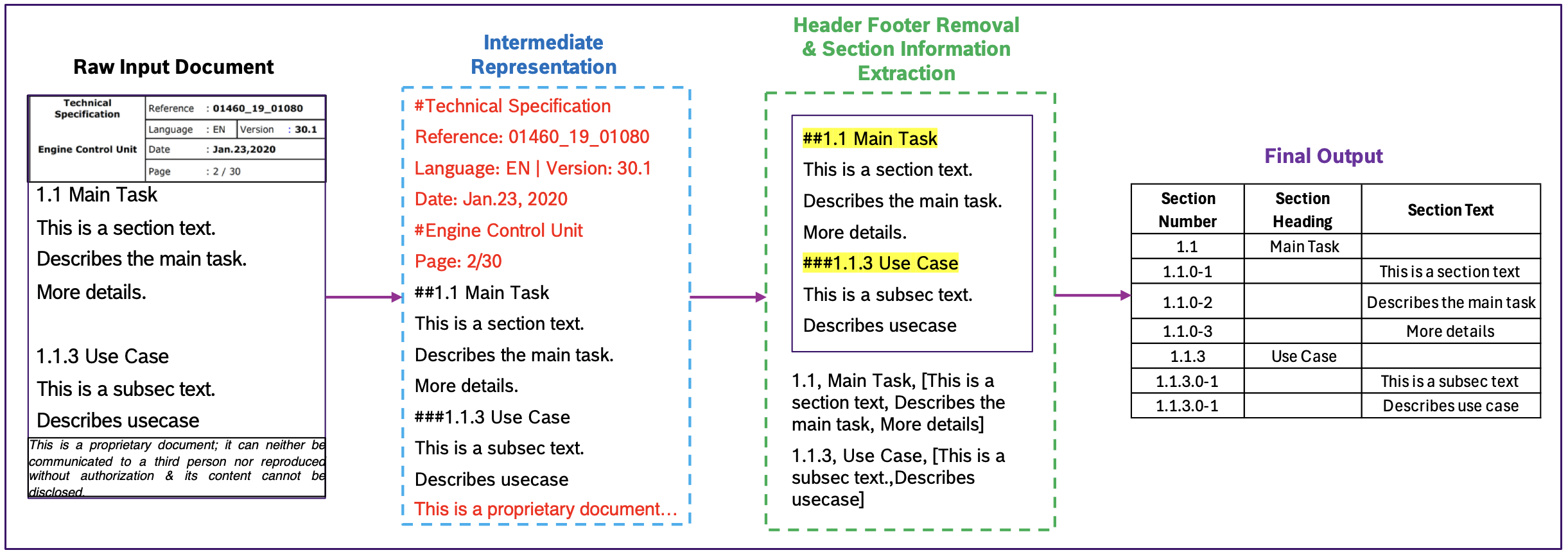}

\end{figure*}

As shown in Figure~\ref{fig:extraction_workflow}, the input document is reduced to sentences with titles prepended with "\#" tags.  While both the approaches do not detect figures, the figure caption gets detected. We preserve the image caption as a placeholder. Better table and image handling is a part of future work.

$\bullet$ \textbf{Header-Footer Removal}: The raw customer requirement documents include header and footer sections, as shown in Figure~\ref{fig:extraction_workflow}. To produce a clean extracted output, it is essential to eliminate this information. We explore bounding box algorithms \cite{d2l_bounding_box} for this purpose but find that the required hyper-parameters vary across different customer documents and between landscape and portrait page orientations.

\begin{table}
\centering
\caption{The Header-Footer removal module performance using Random Forest classifier trained on $3,773$ sentences labeled as \textit{Header-Footer} or \textit{Req.Text}.}
\label{tab:hf}
\resizebox{\linewidth}{!}{
\begin{tabular}{|c|ccc|l|ccc|}
\hline
\multirow{2}{*}{\textbf{label}} & \multicolumn{3}{c|}{\textbf{Dataset Statistics}} &  & \multicolumn{3}{c|}{\textbf{Classification Metrics}} \\ \cline{2-8} 
 & \multicolumn{1}{c|}{\textbf{Train}} & \multicolumn{1}{c|}{\textbf{Test}} & \textbf{Total} &  & \multicolumn{1}{c|}{\textbf{precision}} & \multicolumn{1}{c|}{\textbf{recall}} & \textbf{f1-score} \\ \hline
\begin{tabular}[c]{@{}c@{}}Header-\\ Footer\end{tabular} & \multicolumn{1}{c|}{1146} & \multicolumn{1}{c|}{283} & \begin{tabular}[c]{@{}c@{}}1429\\ (38\%)\end{tabular} &  & \multicolumn{1}{c|}{0.91} & \multicolumn{1}{c|}{0.8} & 0.85 \\ \hline
Req. Text & \multicolumn{1}{c|}{1872} & \multicolumn{1}{c|}{472} & \begin{tabular}[c]{@{}c@{}}2344\\ (62\%)\end{tabular} &  & \multicolumn{1}{c|}{0.89} & \multicolumn{1}{c|}{0.95} & 0.92 \\ \hline
\end{tabular}
}
\end{table}

In the intermediate text representation, headers and footers appear as sentences (highlighted in red in Fig.\ref{fig:extraction_workflow}). We approach this as a binary classification problem, labeling each sentence as either header-footer or requirement text. A lightweight Random Forest classifier \cite{random_forest} is developed using two features: $frequency$ and $position$. We hypothesize that header and footer texts are redundant, occurring multiple times, and have fixed positions—headers at the top and footers at the bottom of pages. We annotate a sample of three documents with $3,773$ sentences into the classes \textit{Header-Footer} or \textit{Req.Text}. The dataset statistics and classifier performance are detailed in Table~\ref{tab:hf}. This trained model is then employed to detect and remove headers and footers.\\




$\bullet$ \textbf{Section Information Extraction}: 
Given a raw customer requirement, we first generate its intermediate text representation. We then eliminate the header and footer to obtain the final requirement text, which is parsed to extract section titles (section number and section heading) and section text. Using markdown text as the intermediate representation, sentences starting with '\#' and containing one or more consecutive instances are labeled as section titles (highlighted in yellow in Fig.~\ref{fig:extraction_workflow}). From the section title, we can easily parse the section number and heading. The sentences between any two section titles $s_i$ and $s_j$ as the section text of $s_i$. The final output is a list of tuples $\left< number_i, heading_i, [text_{i_{1}}, text_{i_{2}} ... , text_{i_{m}}] \right >$


$\bullet$ \textbf{Final Output Generation}: After obtaining the section information as a list of tuples, it is then arranged in a tabular format as shown in Fig.~\ref{fig:extraction_workflow}. The section texts are assigned with the section numbers, that extend the number of the section to which it belongs. 

\label{sec:extraction}
\subsection{Classification}
This module performs the Requirement Type classification task~\cite{seclass, seclass2} on the templatized document from the Extraction module into predefined requirement type categories -- \textit{Info, Header, Functional Requirement} and \textit{Non-Functional Requirement} as shown in Figure~\ref{fig:class}. We make use of the adaptive fine-tuning~\cite{stollenwerk2022adaptivefinetuningtransformerbasedlanguage} of transformer-based language models~\cite{10.5555/3295222.3295349} for requirement text classification. 

\begin{figure*}[thb]
  \centering
  \caption{Requirement Classification using Adaptive Finetuning. Input is requirement documents with/without class labels. Larger chunk of domain-relevant requirement documents used for extended pretraining using masked language modeling. Smaller chunk with class labels used for task aware finetuning for requirement type classification. }
  \includegraphics[width=0.8\linewidth]{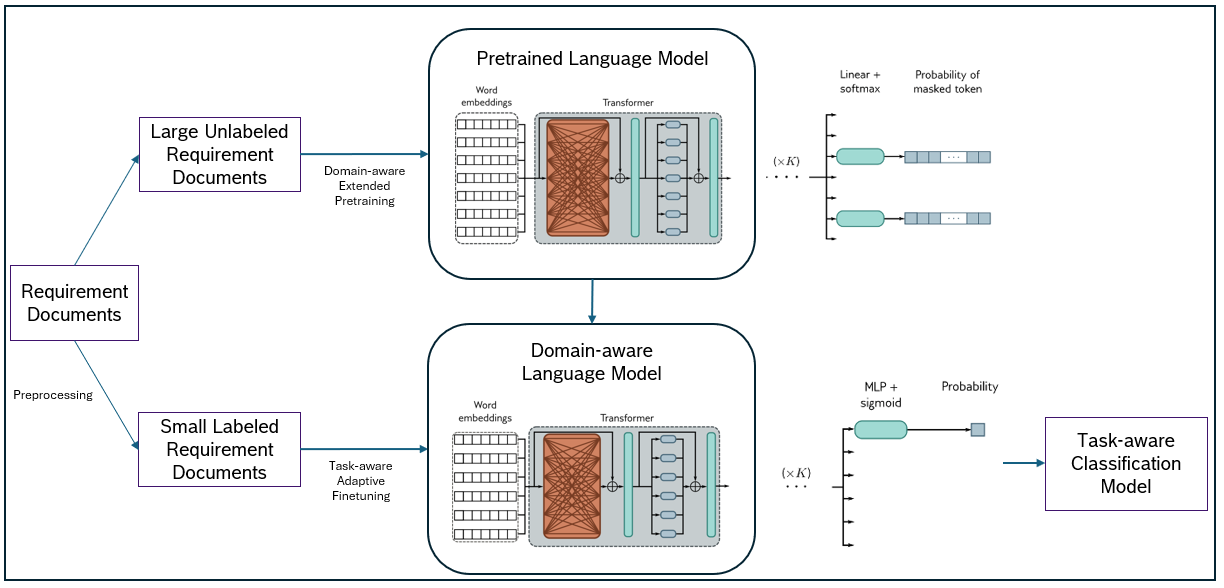}
  \label{fig:classification}

\end{figure*}


The classification pipeline starts with an adaptive fine-tuning phase designed to specialize pre-trained transformer models, such as BERT \cite{devlin2019bertpretrainingdeepbidirectional}, for domain-specific contexts. This process involves performing an additional fine-tuning step on domain-relevant unsupervised corpora prior to task-specific supervised fine-tuning. Thus, it integrates large quantities of unlabeled domain-relevant text with limited annotated data to enhance the model with specialized knowledge. This addresses the challenges posed by task-specific datasets that are out-of-distribution relative to the pre-training data by introducing extra data closer to the distribution of the target dataset, ensuring better alignment with the task at hand. thereby enhancing the model’s ability to handle the domain-specific nuances.

\textbf{Unsupervised Domain Adaptation}: The proposed method, depicted in Figure \ref{fig:classification}) involves a pre-processing pipeline that converts all text to lowercase and removes punctuation, except for structural identifiers like underscores, while retaining critical negations and modal verbs (e.g.,``not'',``shouldn't'') and filtering out conventional stopwords.  This is followed by Iterative Masked Language Modeling \cite{devlin2019bertpretrainingdeepbidirectional} which fine-tunes the pre-trained model on unlabeled domain-specific corpora by masking random tokens and predicting them, enhancing the model's understanding of domain-specific semantics and syntactic structures.

\textbf{Task-Specific Supervised Fine-Tuning}: Building upon the domain-adapted model, supervised fine-tuning is performed using labeled datasets that map requirement sections to predefined classes (ref Figure \ref{fig:classification}). The model architecture extends the base transformer with a classification head that includes a multi-layer perceptron network. The [CLS] token, which encapsulates the document-level semantic representation from the transformer’s output, serves as the input to the network.

\textbf{Requirement Type Classification:} This task evaluates the ability of the model to categorize sections of requirement documents. In this work, we fine-tune a language model on labeled requirement datasets. The goal is to enhance the model's ability to recognize domain-specific terminology, distinguish closely related categories like Functional and Non-Functional Requirements, and generalize across diverse automotive documents.


\label{sec:classification}

\section{Results and Analysis}
In this section, we discuss the performance of the extraction and classification modules. We perform a manual evaluation for the extraction module and automatic evaluation of the classification module. 

\subsection{Extraction}

The gold standard documents for the extraction module are historical data manually curated by domain experts, revised for many iterations in discussion with customers, developers etc. This data therefore contains manually paraphrased section headings and text and additional details that do not directly correspond to the raw customer document. To alleviate this, domain experts suggested to manually evaluate the extraction output themselves. Randomly selected 5 documents were provided to 3 domain experts. The experts then evaluated the overall extraction component and the header-footer classification model independently. 

\noindent
\textbf{Metrics}: We use two metrics used for validating the extraction module of ReXCL -- Overall Evaluation (out of 5), that evaluates the overall extraction quality and Header-Footer Accuracy (in \%) that evaluates the performance of the ML model trained for Header-Footer detection.\\

\noindent \textbf{Overall Evaluation}: For every document, each \textit{row} in the final extraction output  was rated on a score Likert scale of 0-5 independently by the 3 annotators. The final score for a particular document was then an average of all the scores assigned to each row. The score semantics were also described by the experts, which served as an annotation guideline: 

Score 1: Very Poor - Extraction results are significantly inaccurate, with major errors ; Score 2: Poor - Some essential content is extracted incorrectly, leading some inefficiencies ; Score 3: Average - General extraction accuracy is acceptable, but there are notable errors that require occasional corrections; Score 4: Good - The extraction is largely accurate, with only minor mistakes that do not significantly impact workflow; Score 5: Excellent - Extraction is highly accurate, with negligible or no errors, and meets the expectations set for production-level quality.\\

\noindent \textbf{Header-Footer Accuracy:} For the same 5 documents, the experts also give a binary score indicating if a sentence detected as header-footer by the ML model is correct (1) or not (0). We calculate the accuracy between the ML model results and the expert scores.
\begin{table}[!thb]
\caption{Manual Evaluation of the Extraction module}
\label{tab:extr_eval}
\resizebox{\linewidth}{!}{
\begin{tabular}{|c|c|c|c|c|c|c|c|}
\hline
\textbf{Expert} & \textbf{Task} & \textbf{Doc1} & \textbf{Doc2} & \textbf{Doc3} & \textbf{Doc4} & \textbf{Doc5} & \textbf{Average} \\ \hline \hline
\multirow{2}{*}{\textbf{E1}} & Overall (/5) & 4.38 & 4.33 & 4.44 & 4.49 & 4.56 & 4.44 \\ \cline{2-8} 
 & H-F Accuracy (\%) & 100 & 98.55 & 88.83 & 92.86 & 100 & 96.05 \\ \hline \hline
\multirow{2}{*}{\textbf{E2}} & Overall (/5) & 4.33 & 4.38 & 4.43 & 4.5 & 4.56 & 4.44 \\ \cline{2-8} 
 & H-F Accuracy (\%) & 100 & 98.55 & 88.83 & 92.86 & 100 & 96.05 \\ \hline \hline
\multirow{2}{*}{\textbf{E3}} & Overall (/5) & 4.28 & 4.29 & 4.5 & 4.44 & 4.88 & 4.48 \\ \cline{2-8} 
 & H-F Accuracy (\%) & 100 & 98.55 & 88.83 & 92.86 & 100 & 96.05 \\ \hline \hline
\multirow{2}{*}{\textbf{Average}} & Overall (/5) & 4.33 & 4.33 & 4.28 & 4.48 & 4.67 & 4.42 \\ \cline{2-8} 
 & H-F Accuracy (\%) & 100 & 98.55 & 88.83 & 92.86 & 100 & 96.05 \\ \hline
\end{tabular}
}
\end{table}

\noindent
\textbf{Analysis:} Table~\ref{tab:extr_eval} shows the expert scores for 5 documents for both the Overall evaluation and the Header-Footer (H-F) accuracy. The IAA between the experts for the overall evaluation metric as measured by Pearson Correlation (as the scores were was 0.92, indicating high levels of agreement. We find that on average the experts feel that the extraction quality is good, thus obtaining a score of 4.4 out of 5. 

On inspecting the scores for the documents, we find two major errors that the extraction model makes: (i)~the first few pages of the document comprising of the title page, table of contents, mixed with header footer variations, author information etc. are in a heterogeneous format, making it difficult for the intermediate text representation module to accurately identify the textual units to output. (ii)~sometimes the section texts are "No Requirement", "Not Applicable", "N.A." etc., which appears multiple times across different sections. The header-footer model incorrectly classifies them as header-footer texts and remove them. Hence, the section texts are lost.

For the Header-Footer accuracy, we find that the experts highly align on the score, since it is less subjective compared to the overall evaluation metric. The overall accuracy is 96\%, thus showing the strength of simple ML models. Note that, the classifier was trained on two features -- frequency and position -- both of which are language agnostic. The errors made by the classifier are mainly misclassifying texts like "No Requirement", "Not Applicable", "N.A." as discussed above.  

\begin{figure}[!thb]
  \centering
  \caption{Heatmap of annotator scores on a sacle of 0-5}
  \includegraphics[width=\linewidth]{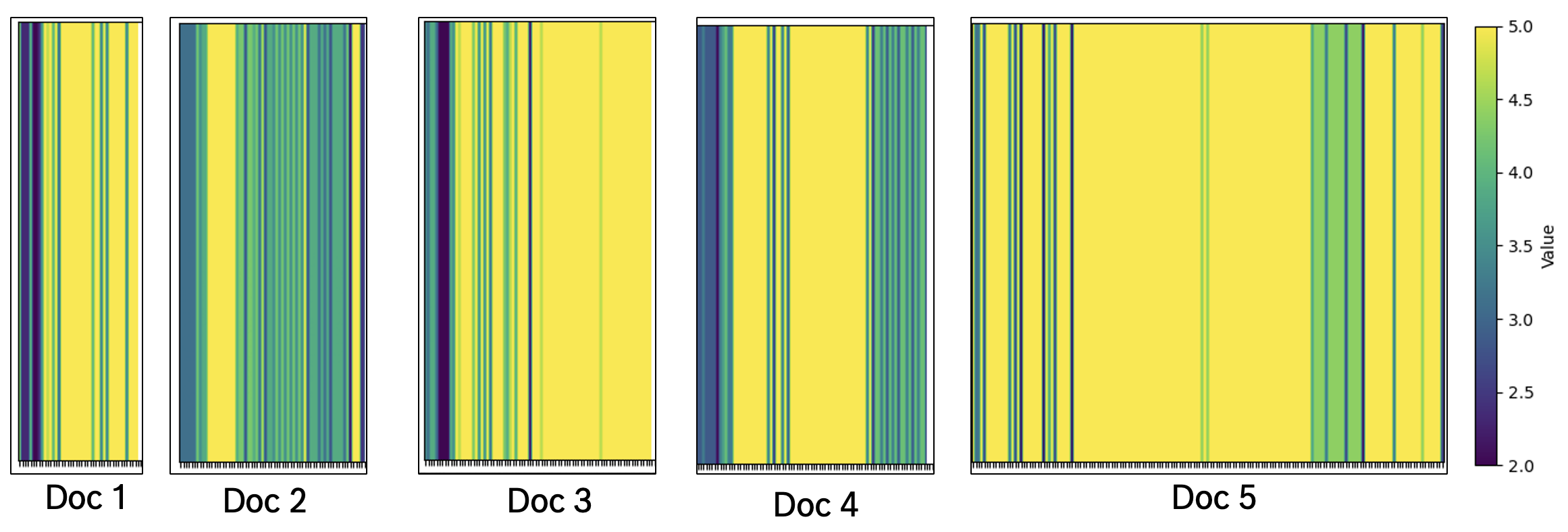}
  \label{fig:heatmap}
  \vspace{-5mm}
\end{figure}

Figure~\ref{fig:heatmap} shows the heatmap over the five documents according to the scores assigned during manual evaluation. We find the method performs reasonably well in most parts of the document. However, it faces challenges in the initial parts of the document. This is because the initial parts of the document contain information in heterogenous formats like customer information in tables and figures, authors, history, table of contents etc. This leads to confusion in parsing front page headers where headers and text often get mixed up, along with the detection of watermarks which further complicates parsing. 

Additionally, table and image extraction need significant improvements to ensure accuracy and consistency. Another major issue is the handling of multiple requirements within a single block, which leads to ambiguity and misinterpretation. Furthermore, the mixing of tables with headers and footers disrupts the document structure, making it difficult to extract and organize information correctly.
\label{sec:eval_extraction}

\subsection{Classification}
In this section, we evaluate the effectiveness of the adaptive fine-tuning approach applied to the classification task outlined in Section 2.2. The adaptive fine-tuning process utilizes a dataset consisting of approximately $270k$ requirement texts. These texts were meticulously extracted and curated by domain experts from customer requirement documents. For the classification task, we employed a curated dataset of requirement texts that have been manually labeled with their respective Requirement Types.
The data statistics are given in the Table \ref{tab:dataset_statistics}. We employed the pretrained multilingual BERT model \cite{bert_base_multilingual_uncased} for our classification tasks, along with a fine-tuned version that utilized the Adaptive Fine-tuning approach

\begin{table}[t]
\small
    \caption{Dataset statistics for the Classification task}
    \label{tab:dataset_statistics}
    \centering
    \begin{tabular}{|>{\raggedright\arraybackslash}p{2.6cm}|>{\centering\arraybackslash}p{1cm}|>{\centering\arraybackslash}p{1cm}|>{\centering\arraybackslash}p{1cm}|}
        \hline
        \multirow{2}{*}{\textbf{Task}} & \multicolumn{3}{c|}{\textbf{Statistics}} \\ \cline{2-4}
                                          & \textbf{Train} & \textbf{Test} & \textbf{Total} \\ \hline
        Adaptive Fine-tuning & 230,059 & 40,598 & 270,658 \\ \hline
        Requirement Classification & 6,564 & 2,813 & 9,377 \\ \hline
    \end{tabular}
\end{table}

\begin{table}[!thb]
\small
    \caption{The results for the Classification task}
    \label{tab:classification}
    \centering
    \begin{tabular}{|l|>{\centering\arraybackslash}p{1.5cm}|>{\centering\arraybackslash}p{1.5cm}|}
        \hline
        \multirow{2}{*}{\textbf{Class Label}} & \multicolumn{2}{c|}{\textbf{Classification F-Score}} \\ \cline{2-3}
                                          & Vanilla BERT & Adaptive Finetuned BERT                     \\ \hline
        HEADER            & 0.41                          & \textbf{0.99}                        \\ \hline
INFO              & 0.40                           & \textbf{0.98}                       \\ \hline
FUNC\_REQ         & 0.24                            & \textbf{0.98}                        \\ \hline
NON\_FUNC\_REQ    & 0.22                         & \textbf{0.93}                        \\ \hline
    \end{tabular}
\end{table}


\textbf{Analysis:} The experimental results, presented in Table \ref{tab:classification}, demonstrate that incorporating the adaptive fine-tuning phase with the BERT model substantially enhances classification performance when compared to the vanilla BERT model, which is solely pretrained on open-source text documents. The MLM-based unsupervised fine-tuning step enhances the BERT model’s understanding of domain-specific features of requirement documents, which translates into improved contextual embeddings. These embeddings enable the BERT model to better differentiate between subtle distinctions inherent to requirement sections, such as those between informational content and actionable requirements. Also, under-represented class label such as Non-functional requirement can be classified with much better accuracy using this method. Thus, adaptive fine-tuning strategy ensures that  model maintains its domain adaptability while excelling in task-specific requirement classification.

\label{sec:eval_classification}
\label{sec:eval}

\section{Deployment} 
\begin{figure*}
\centering
\begin{subfigure}{.5\textwidth}
  \centering
  \includegraphics[width=.95\linewidth]{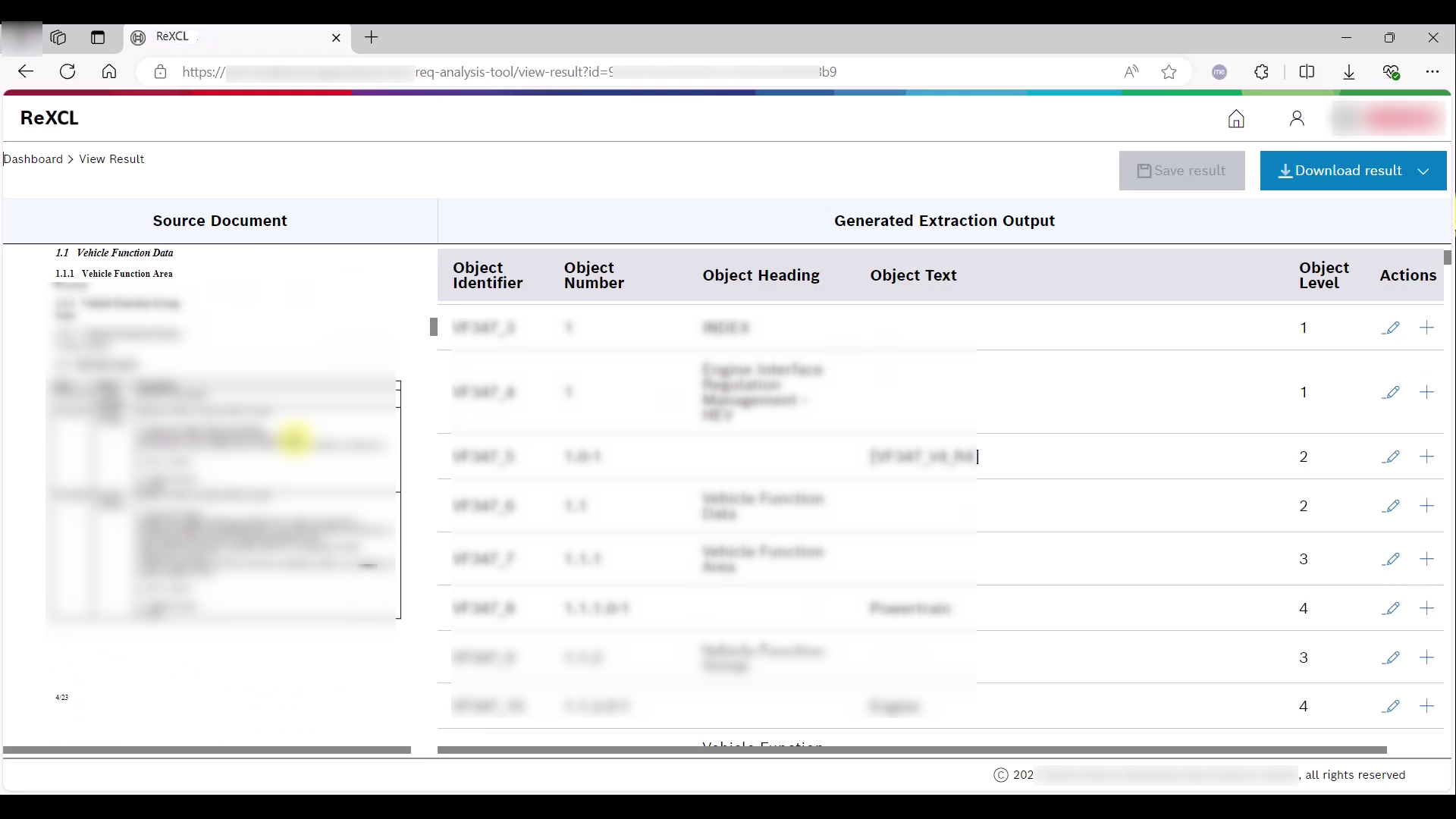}
  \caption{ReXCL Tool - Extraction View }
  \label{fig:ext}
\end{subfigure}%
\begin{subfigure}{.5\textwidth}
  \centering
  \includegraphics[width=.95\linewidth]{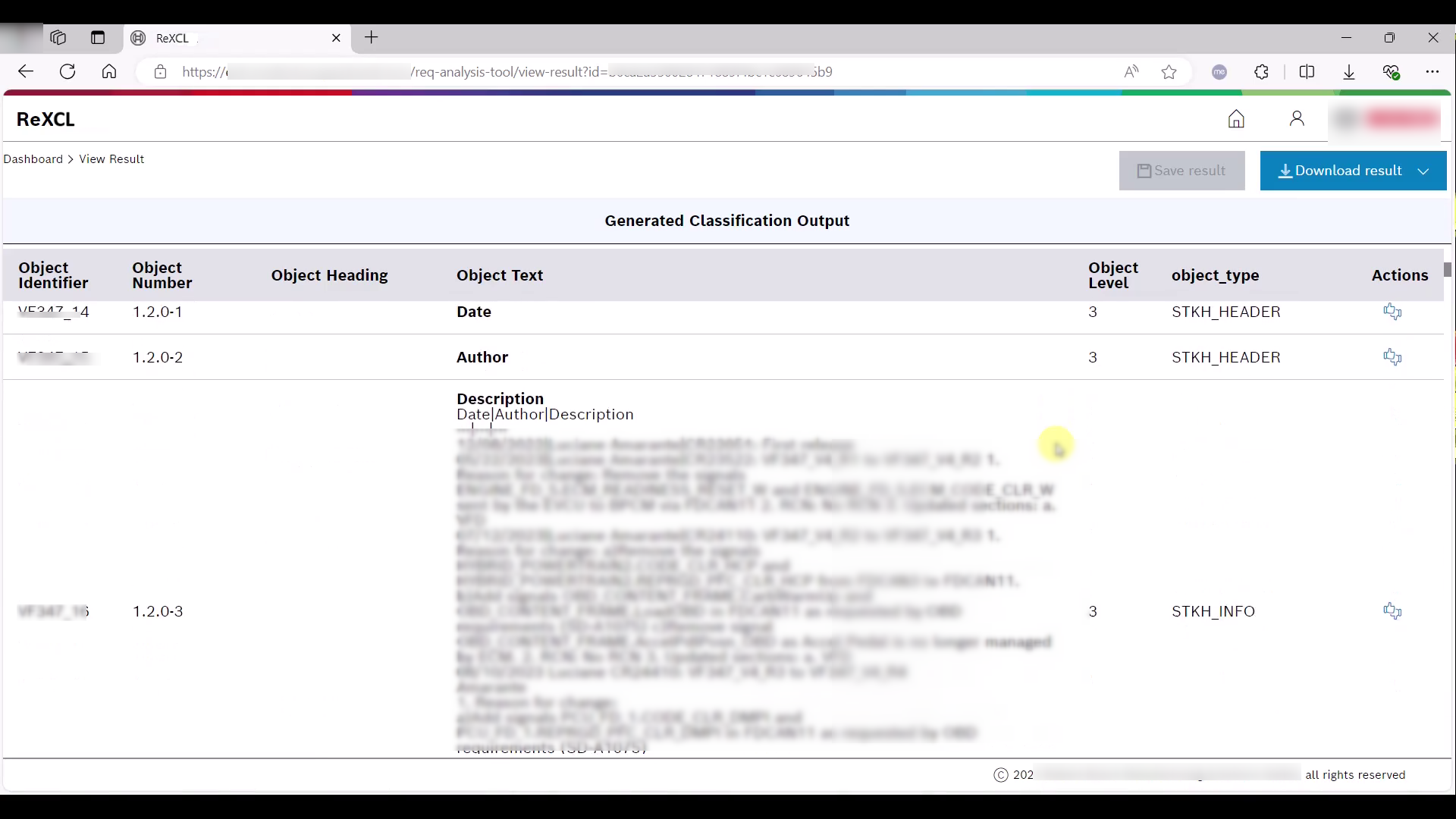}
  \caption{\textcolor{black}{ReXCL Tool - Classification View}}
  \label{fig:class}
\end{subfigure}
\caption{ReXCL Tool Overview - Requirement document extracted in the structured format from word documents/PDF. Then the extracted texts from requirement document classified into requirement types : Header, Info, Functional and Non-Functional requirements.}
\label{fig:test}
\end{figure*}
The tool for requirement document extraction and classification has been deployed and is actively being used inside the organization. The frontend has been developed using Angular, the backend is python FastApi and the database used to store the documents and the intermediate results is MongoDB. The extraction module works comfortably in a CPU while the classification module requires a GPU. The tool therefore is currently deployed in a GPU server of 6GB RAM. 

Figures ~\ref{fig:ext} and ~\ref{fig:class} shows the screenshots of the deployed tool. For security reasons, the sensitive information has been blurred. The left pane of Fig. ~\ref{fig:ext} shows the original requirement document (in pdf). The right hand pane shows the structurized information containing Object Identifier, Object Number (derived from the section numbers), Object Heading (derived from the section headings), Object Text (the corresponding text associated in the section) and Object Level (derived from the object number column using heuristics). The header and footers do not appear in the output. Also we provide the "Actions" tab where the user can modify any incorrect output generated. 

The "Download" button on the top-right enables the user to download the result in different formats like csv, excel, yaml and json. This output can be used independently by the user or he/she can proceed to the "Classification" module shown in Figure~\ref{fig:class}. Here each Object text is labelled using an Object Type which is one of the labels mentioned in Section~\ref{sec:classification}. The feedback is captured through "Action" where the user can mention if the classifier output is correct or incorrect. In case the label is incorrect, the tool prompts the user to provide the correct label which is then saved. Similar to the extraction module, the results for classification can be downloaded in a structured format, which now contains the "Object Type" information.
\label{sec:deploy}

\section{Conclusion}

The ReXCL tool represents a significant advancement in the automation of extraction and classification processes within requirement engineering. By effectively processing raw documents and categorizing requirements into distinct classes, ReXCL enhances the efficiency and accuracy of managing software requirements. The integration of heuristics and predictive modeling, along with adaptive fine-tuning of models like BERT, demonstrates its robust capabilities. Our evaluations confirm the tool's effectiveness, paving the way for improved practices in requirement management. 


In this study, we have not examined effective handling of images and tables. As part of our future work, we plan to enhance ReXCL to efficiently manage heterogeneous and multi-modal data, including presentations and Excel documents.

\bibliography{references}
\bibliographystyle{acl_natbib}




\end{document}